\documentclass[twocolumn,showpacs,preprintnumbers,amsmath,amssymb]{revtex4}
\usepackage[dvips]{graphicx}% Include figure files
\usepackage{dcolumn}% Align table columns on decimal point
\usepackage{bm}% bold math
\usepackage{color} % load color package
\begin{document}

\preprint{APS/123-QED}
\title{The Effect of Transfer Printing on Pentacene Thin-Film Crystal Structure}% Force line breaks with \\
\author{Y. Shao }
\author{S. A. Solin }%
\altaffiliation{Fax: 314-935-5983\\}%Lines break automatically or can be forced with \\ 
\email{solin@wustl.edu}
\affiliation{%
Washington University in St. Louis, Center of Materials Innovation, 
Department of Physics, St. Louis, Missouri 63130\\}%
\author{D. R. Hines and E. D. Williamsr}
\affiliation{
University of Maryland, Department of Physics and Laboratory for Physical Sciences, College Park, Maryland 20742\\}%

\date{\today}% It is always \today, today,

\begin{abstract}

The thermal deposition and transfer Printing method had been used to produce pentacene thin-films on SiO$_{2}$/Si
 and plastic substrates (PMMA and PVP), respectively. X-ray diffraction patterns of pentacene thin films showed 
 reflections associated with highly ordered polycrystalline films and a coexistence of two polymorphs phases classified 
 by their d-spacing, {\it{d(001)}}: 14.4 and 15.4 $\AA$ .The dependence of the c-axis correlation length and the phase  fraction 
 on the film thickness and printing temperature were measured. A transition from the 15.4  $\AA$ phase towards 14.4  $\AA$ 
 phase was also observed with increasing film thickness.  An increase in the c-axis correlation length of approximately 
 12$\sim $16\%  was observed for Pn films transfer printed onto a PMMA coated PET substrate at 100$\sim$ 120 $^{o}$C as compared to 
 as-grown Pn films on SiO$_{2}$/Si substrates. The transfer printing method is shown to be an attractive for the fabrication 
 of pentacene thin-film transistors on flexible substrates partly because of the resulting improvement in the quality of the 
 pentacene film. 
\end{abstract}

\maketitle

\section{\label{sec:level1}Introduction\protect\\}

Pentacene (C$_{22}$H$_{14}$), a linear acenic hydrocarbon composed of five benzene rings[1,2], has received 
considerable interest as the active layer for organic thin-film transistors (TFTs). This is due to the
 strong tendency of pentacene (Pn) to form an ordered film that exhibits superior transport properties[3,4]. 
 To make organic TFTs with high mobility, it is very important to fabricate highly ordered pentacene
   films with large sized crystal domains[5], and low defect density[6]. Several fabrication methods, for example
    thermal evaporation[7, 8,9,10], growth from solution[10,11], have been used to grow Pn on various non-organic 
    substrates, such as SiO$_{2}$ [9,12], carbon[13], Kapton[10,14] and NaCl[17]. Previous structure studies by X-Ray diffraction
     (XRD)[10,11,12,17] and electron diffraction (ED)[10,15] found at least four polymorphs of Pn. They are classified by
      their layer spacing,{ \it{d(001)}}[10] : 14.1[11,16], 14.4[1,2], 15.1[17], and 15.4  $\AA$[9, 12,18]. 
      The 14.1  $\AA$ structure was
       commonly referred to as the Òsingle crystal phaseÓ. The 15.1 and 15.4  $\AA$ were referred to as the 
       Òthin film phaseÓ. The 14.4  $\AA$ phase was first identified by  Campbell et al.[1,2] by using a film method. 
       It was also called a Òsingle crystal phaseÓ[5, 8 ,9, 12, 18, 20] in the past. However Mattheus et al.[10] showed that this 
       phase should be designated as  Ò thin film phaseÓ. The unit cell parameters [10, 15,17,21] and the stacking of the molecules 
       within the layer, which affect the electronic properties of Pn[6, 22], were also  studied.
       
Currently, research efforts addressing pentacene are being driven by an interest in developing and improving high-quality organic thin-film 
devices for low cost fabrication on flexible organic substrates[19]. Hines et al. have recently reported that an adaptation 
of Nano-imprint lithography (NIL) [23, 24]  called Nano-Transfer Printing[ 25],  can be used to create high-quality organic
 TFTs on flexible substrates. The pentacene TFTs exhibited a field-effect mobility as high as 0.09 cm$^{2}$ (Vs)$^{-1} $and an
  on/off ratio of 10$^{4}$ [25].
  
In this paper, we present XRD studies of pentacene films deposited on SiO$_{2}$/Si and printed onto PMMA and PVP
 substrates as a function of film thickness, printing temperature, and printing time. We have observed the co-existence of the 14.4 and 15.4  $\AA$
 phases And an increase in the c-axis correlation length (i.e. Pn film quality perpendicular to the substrate surface) 
 introduced by the transfer printing. 
\section{Experiments}
Pentacene films with different film thicknesses deposited on (SiO$_{2}$)/Si wafers were fabricated by thermal deposition in a high vacuum
 evaporation chamber.  The Si$\langle$001$\rangle$ wafers with a 300 nm thermal oxide surface (SiO$_{2}$) were used as a transfer substrate in 
 a subsequent transfer printing process.  The Pn films on a PMMA/PET substrate were fabricated by the transfer printing method. The 200
  nm thick PMMA surface coating was spin coated on PET at 4000 rpm for 60 sec and baked on a 90 $^{o}$C hotplate for 3 minutes. 
  prior to the transfer printing. The PET substrate was a DuPont Molinex@ 453 film, with an ITO coating on the back surface.
   The transfer printing was done with a Nanonex NX2000 imprinter at a fixed printing pressure (100 psi) with different printing 
   control parameters, such as temperature and time. The details of the transfer printing process have been addressed elsewhere[25].
   
The out-of-plane structure of the pentacene films was determined using X-ray diffraction in a coupled $\Theta$-2$\Theta$ reflection geometry. 
Ultra high-resolution XRD studies were performed using a Rigaku UltraX18 rotating anode generator equipped with a Huber 4-circle
 diffractometer system. The x-ray source used was Cu{\it{K}}$\alpha.$ Data acquisition was performed with a uniform X-ray power of 14 kW 
 and a selectable scanning dwell time.  All the measurements were performed at room temperature under ambient conditions.
 
\section{Results and Discussion}
The structural properties of pentacene films were studied as a function of layer thickness and transfer printing temperature.  
Figure 1 shows the XRD patterns of 0.05, 0.25, 0.5 and 1 $\mu$m  pentacene thin-films deposited on SiO$_{2}$ /Si substrates prior to transfer printing.
 The reflection intensities are plotted vs. the scattering vector q ($\AA$$^{-1}$), where 
\begin{eqnarray}
\
q=\frac{4\pi\sin(\Theta)}{\lambda}[26]
\label{eq:one}.
\end{eqnarray}
The d-spacing was calculated from the inverse of the slop in a q-plot [27]. 

\begin{figure} 
 \centering 
 \includegraphics[width=0.5\textwidth]{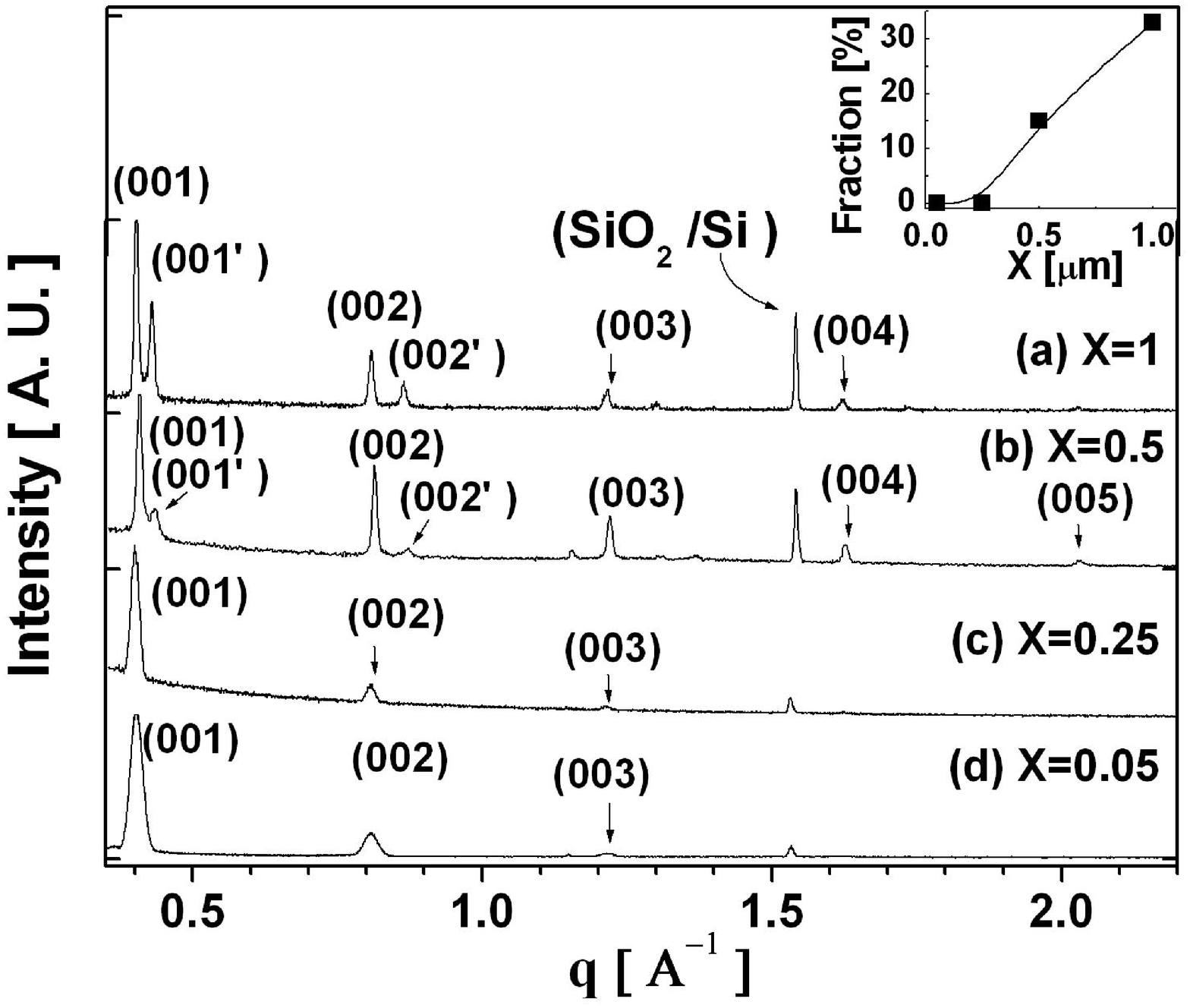} 
 \caption{Figure 1. XRD patterns of pentacene thin films deposited on SiO$_{2}$/Si 
 substrates prior to transfer printing with different film thickness x $\mu$m. (a) x=1, (b) x=0.5, (c) x=0.25, and (d) x=0.05. 
  The inset shows the volume fraction of 14.4 $\AA$phase as a function of film thickness.} 
 \label{...} 
\end{figure}

 In figures 1(a)-(d), a set of sharp reflections
 associated with the Pn film are present in these x-ray patterns and can be indexed as{ \it{(00n)}} reflections, with the{ \it{(001)} }
 plane d-spacing equal to 15.40$\pm$0.4 $\AA$. Also in Figs. 1(a)\&(b), a second set of sharp reflections are visible in the 0.5
  and 1.0 $\mu$ m thick films that can be indexed as {\it{(00n')}} reflections with a {\it{(001')}} plane d-spacing of 14.40$\pm$0.03  $\AA$. 
  The first time observation of the two-phase coexistence was by Dimitrakopoulos[18] who used a SiO$_{2}$ substrate. Additional experiments showed 
  that the 14.4 and 15.4 $\AA$ phases are substrate-induced[10]. They are commonly found in Pn films onSiO$_{2}$ substrate[9, 
  12] and with large film thickness[5, 10, 18] and high growth temperature[5, 10]. With an increase in Pn film thickness, 
  the intensity of reflections associated with the 15.4 $\AA$ phase decreases and the intensity of reflections associated with 
  the 14.4 $\AA$ phase increases. In Fig. 1 inset (as grown Pn on SiO$_{2}$/Si), the volume fraction of the 14.4 $\AA$ phase increased 
  from 0 \% for the 0.05 $\mu$m and 0.25 $\mu$ films to 15\% for the 0.5 $\mu$m film and 33\% for the 1.0  $\mu$m film. This observation agrees
  with previous results[5, 10, 12, 18].The volume fraction here is defined as the ratio of relative intensity of the 14.4 $\AA$ phase
   to the total intensity for the{ \it{(001) }}reflections[19].
   
Figures 2(a)-(e) show the XRD patterns of 0.5  $\mu$m Pn thin films before and after having been transfer printed onto a 
PMMA coated PET substrate with printing conditions of 100 psi at 100, 120 and 140 $^{o}$C for 3 min. and onto a PVP
 coated PET substrate with printing temperature 140  $^{o}$C. 
 
 \begin{figure} 
 \centering 
 \includegraphics[width=0.5\textwidth]{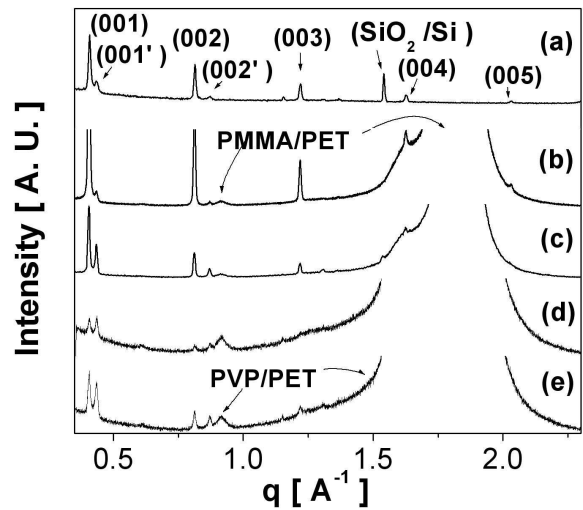} 
 \caption{Figure 2(a). XRD patterns of as-printed 0.5 $\mu$m pentacene thin films on SiO$_{2}$/Si substrate. 
 (b) XRD patterns of pentacene printed on PMMA/PET substrates at 100 psi and 100 $^{o}$C for 3 min. (c) 
 Printed on PMMA/PET at 120 $^{o}$C. (d) Printed on PMMA/PET at 140 $^{o}$C. (e) Printed on PVP/PET at 140 $^{o}$C.} 
 \label{...} 
\end{figure}
 
 The two phase co-existence of the14.4 and 15.4 $\AA$ phases persisted
 through the transfer printing process. The unchanged {\it{d(001)} }spacing indicates that the transfer printing did not affect the 
 rotation of pentacene molecules along {\it{[001]}} direction[17, 21]. A decrease in intensity of the reflections associated 
 with the 15.4 $\AA$ phase and an increase in intensity of the reflections associated with the 14.4 $\AA$ phase are visible 
 with increasing printing temperature.  At 100, 120, and 140 $^{o}$C, the volume fractions of the 14.4 $\AA$ phase on the PMMA substrate are 5, 30, 
 and 55\% respectively. At 140 $^{o}$C, the volume fraction of the 14.4 $\AA$ phase on the PVP/PET substrate is 45\%. 
 Figure 3 shows the expanded XRD patterns of the first and second reflections of 0.05 $\mu$m and 0.5  $\mu$m pentacene films 
 on a PMMA substrate.  Clearly, the 14.4 $\AA$ phase volume diffraction depended on the nature of the substrate, film 
 thickness and printing temperature. It is more thermally stable than 15.4 $\AA$ phase.
 \begin{figure} 
 \centering 
 \includegraphics[width=0.55\textwidth]{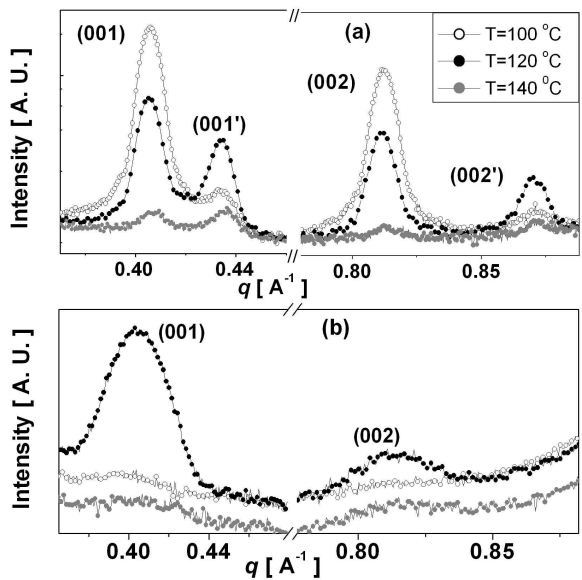} 
 \caption{Figure 3. (a) The first and second reflections of 0.5 $\mu$m pentacene thin films printed on PMMA/PET substrates 
 at different printing temperature. (b) The first and second reflections of 0.05 $\mu$m pentacene thin films printed on PMMA/PET 
 substrates at different printing temperatures.} 
 \label{...} 
\end{figure}
 
To study the crystal perfection along the growth direction (c-axis), we calculated the crystalline grain size of the
 Pn films perpendicular to the substrate surface using Para-crystal theory[28].  Since the (00n) planes are parallel 
 to the substrate, this theory can provide a calculation of the average grain size (c-axis correlation length) in the
  direction normal to the substrate surface.  The broadering equation is given by
\begin{eqnarray}
(\delta s)^{2}=(\delta s)^{2}_{c}+(\delta s)^{2}_{II}=\frac{1}{\bar{L}_{hkl}^{2}}+\frac{(\pi g_{_{II}})^{4}}{d^{2}_{hkl}}n^{4},
\end{eqnarray}
where
\[\delta s =\frac{2\cos\Theta\delta\Theta}{\lambda} \]
is the overall broadening of the n$^{th }$x-ray reflection. ($\delta$s)$_{c}$ is the broadening due to the average grain size and ($\delta$s)$_{II} $
is the broadening due to the structural disorder g$_{II}$. Also $\delta$$\Theta$ is the half width of the peak at half maximum as calculated 
from a Gaussian fit. The ($\delta$s)$^{2 }$values are plotted in Fig. 4(a) as a function of n$^{4}$ for the 15.4 $\AA$ phase of 0.05, 0.25, 0.5
 and 1 $\mu $m  pentacene thin-films deposited on SiO$_{2}$/Si (filled data points) and transfer printed onto the PMMA substrate 
 (open data points).  The average crystalline grain size parameters L$_{hkl}$ and g$_{II}$ were calculated from the ordinate
  intercept and the slop of the lines from a fit of Eq. 2 to the data points in Fig. 4(a). The calculated average grain sizes 
  for the Pn films on SiO$_{2}$/Si substrates prior to transfer printing are shown in Fig. 4(b) (filled data points) and are 227$\pm$9
   ($\sim$15 monolayer (ML)), 409$\pm$16 ($\sim$27 ML), 673$\pm$27($\sim$44 ML), and 805$\pm$32 $\AA$ ($\sim$52 ML) for the film thickness of
    0.05, 0.25, 0.5, and 1 $\mu$ m respectively. We didnÕt find any reflections related to 0.25 $\mu$ m pentacene on PMMA/PET 
    with printing conditions of 100 psi at 100$^{o}$C for 3 min. The reason is not clear. It may due to the damage of the sample 
    during transfer printing process.  For comparison, the calculated average grain sizes for the Pn films on PMMA/PET 
    substrates after transfer printing are also shown in Fig. 4(b) (open data points). For the transfer printed films with 
    thickness 0.05, 0.5 and 1 $\mu$m, the corresponded average grain sizes are 263$\pm $11 ($\sim$17 ML), 750$\pm$30 ($\sim$49 ML) and 907$\pm$
    36 $\AA$ ($\sim$59 ML) respectively. For the above calculations, the instrumental broadening (determined from the broadening of 
    (004)) reflection of the single crystal Si) has been subtracted from the x-ray diffraction peak widths. The term g$_{II} $is about 1$\sim$4\% for both the as-grown 
    and transfer printed Pn films which indicates that the structural perfection along the c-axis direction is high and 
    unaffected by the transfer printing.  Figure 4(b) shows that after transfer printing at 100 $^{o}$C, the average crystalline 
    grain sizes of pentacene thin films increase by about 12$\sim$16 \%.
    
     \begin{figure} 
 \centering 
 \includegraphics[width=0.5\textwidth]{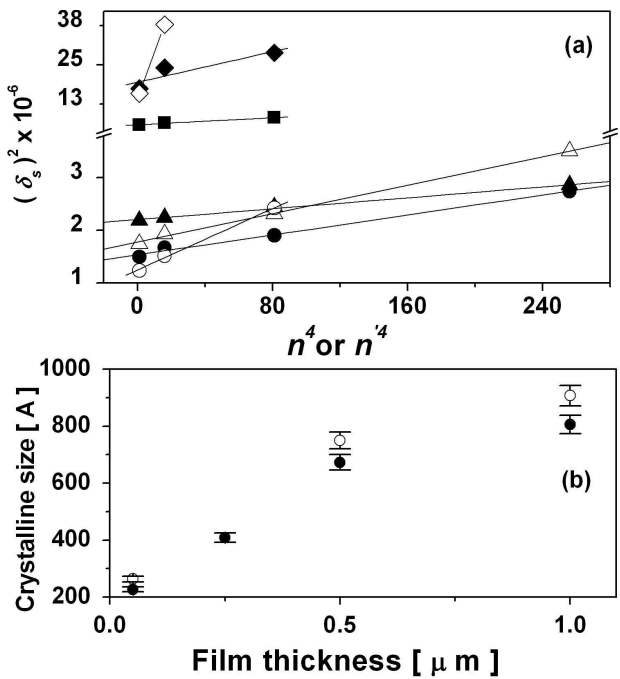} 
 \caption{Figure 4(a) Reflection broadening vs. diffraction order {plotted {as ($\delta$s)$^{2}$ vs. n$^{4}$}} for pentacene films. $\blacklozenge$ 0.05 ; 
$\blacksquare$  0.25; $\blacktriangle $  0.5; $\bullet$ 1$\mu$m as-grown pentacene on SiO$_{2}$/Si; $\Diamond $ 0.05; $\triangle$0.5; $\circ$ 1 $\mu$m 
transfer printed pentacene on PMMA/PET. 
 (b) Calculated average crystalline grain size vs. film thickness for pentacene deposited on SiO2/Si prior to transfer printing 
 (e) and pentacene transfer printed onto PMMA/PET at 100 psi and 100 $^{o}$C for 3 min. (a).} 
 \label{...} 
\end{figure}    
     \begin{figure} 
 \centering 
 \includegraphics[width=0.45\textwidth]{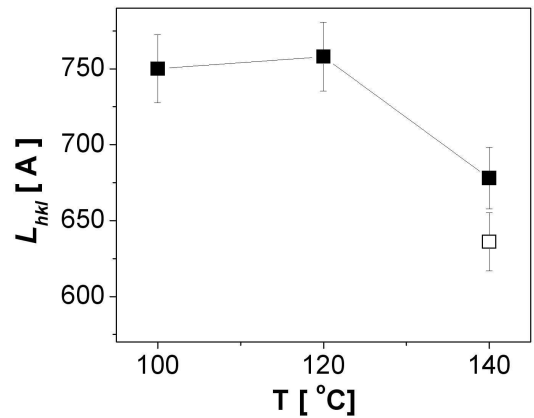} 
 \caption{Figure 5. Crystalline sizes of of 0.5 $\mu$m pentacene thin films as a function of printing temperature. 
$\blacksquare$ Printed on PMMA/PET substrate.  $\Box $ Printed on PVP/PET substrate.} 
 \label{...} 
\end{figure}   
To investigate the influence of the printing temperature on the film crystalline grain size,  0.5 $\mu$m pentacene films were 
printed at temperatures of 100, 120, and 140 $^{o}$C for 3 minutes each. Figure 5 shows the calculated c-axis correlation length 
as a function of printing temperature on PMMA (filled data point) and PVP substrates ( open data point).  Above 120 $^{o}$C 
an increase in printing temperature leads to a decrease of the average crystalline grain size (from 758$\pm$23  at 120 $^{o}$C to 
678$\pm$20  at 140 $^{o}$C).  The crystalline grain size of pentacene printed at 140 $^{o}$C is close to that of the original film on the
 SiO$_{2}$/Si substrate prior to transfer printing but with a higher fraction of bulk crystal phase. 
 
To determine the influence of printing time, pentacene films were printed at 100 $^{o}$C for 3 and 30 minutes. No strong 
evidence showing a change in c-axis correlation length with printing time was observed. 
\section{Conclusions}
In summary, we have studied the structure and calculated the average crystallite grain size of transfer printed pentacene films 
as a function of substrate, film thickness and printing temperature. For small film thickness Pn films showed only 15.4 $\AA$ phase.
 For larger film thickness, the 14.4 $\AA$ phase appeared and coexisted with 15.4 $\AA$ phase. The volume fraction of 
the 14.4 $\AA$ phase increases with increasing film thickness. This trend persisted during the transfer printing. The volume fraction 
of the 14.4 $\AA$ phase increased with printing temperature. Transfer printing at 100 and 120 $^{o}$C improves the c-axis correlation 
length of Pn films printed onto a PMMA coated PET substrate. Longer printing times did not contribute any measurable improvement. 
 Generally, transfer printing a pentacene film from a SiO$_{2}$/Si substrate to a plastic substrate (PMMA/PET) increased the crystalline
  grain size by approximately 12$\sim$16 \% at the lower printing temperatures (100 -120 $^{o}$C). This improvement in the c-axis correlation length 
  of the transfer printed Pn films highlights an important advantage of the transfer printing method and adds to the attractiveness of 
  this technique for the fabrication of high quality pentacene thin films on plastic substrates.

\begin{acknowledgments}
This work was supported in part by NSF Grant No. ECS-0329347 and by the Laboratory for Physical Sciences, University of Maryland.
\end{acknowledgments}

% Produces the bibliography via BibTeX.

\end{document}